# Finding the ET Signal from the Cosmic Noise

**Ross Davis, PhD**
Indiana University
rdavis70@protonmail.com

**April 8, 2022**

## Abstract

This paper highlights a methodological approach designed to enhance the search for extraterrestrial intelligence (SETI) by hypothesizing that a transmission technosignature would likely have two features: 1) be wideband in the microwave or higher frequency range that originates from a hub within a supposed ET interplanetary navigation/communication (nav/comm) network, and 2) contain x-ray pulsar-based navigation (XNAV) metadata.

Potential contributions to the field include improved accuracy in finding transmission technosignatures and other technosignatures in the electromagnetic spectrum, a common standard in reaching a Schelling Point (a mutual realization of how we and ETs can find each other), and operationalizing models such as the Drake Equation.

# 1. Introduction

It has been proposed that the search for extraterrestrial intelligence (SETI) can be done by locating potential extraterrestrial (ET) interplanetary navigation/communication (nav/comm) networks in space (Davis, 2019). In this paper a modification of the aforementioned approach is proposed that considers a communicative ETI potentially using wideband signals and a pulsar positioning system (PPS)—one based on an X-ray pulsar-based navigation (XNAV) solution (Vidal, 2017)—as a component of its interplanetary nav/comm network. This is expressed more specifically through the following two hypotheses:

> H1: A transmission technosignature would more likely be a wideband signal in the microwave or higher frequency range, which originates from the hub of an ET's interplanetary nav/comm network.
>
> H2: A transmission technosignature would likely be a wideband signal in the microwave or higher frequency range, which contains x-ray pulsar-based navigation (XNAV) metadata.

The second section presents a method in which to test the hypotheses, using network analysis on exoplanetary and pulsar data, so as to create visualizations of hypothetical ET interplanetary nav/comm networks that would utilize XNAV. The last section concludes with potential contributions to SETI as an emerging field within mainstream science.

# 2. Concepts

The following defines concepts associated with the approach, which is designed to enhance SETI by hypothesizing that a transmission technosignature would more likely originate from the hub of a supposed ET interplanetary nav/comm network, and that the signal would likely contain XNAV metadata. The concepts include transmission technosignature, ET interplanetary nav/comm networks, and ET XNAV; in addition, two hypotheses that are based on these concepts.

## 2.1. Transmission Technosignatures

By transmission technosignatures (or ET space signals) we mean transmission of information through space on carriers such as photons (including radio and laser signals), which may be intended for us here on Earth, or detected as leakage.

One way a transmission technosignatures can be transmitted or received is directly between a pair of transceivers, with one transceiver in orbit around or located on an exoplanet in one solar system, while the other transceiver is similarly situated in another solar system elsewhere in the network; this is without the use of signal relays in between. A variant of this can be indirectly, through signal relay satellites positioned intermittently along the path between the transceivers, whereby the satellites help to direct and boost a signal over interstellar distances. Such satellites can be artificial (made by the ET civilization) or natural, such as using the gravitational lensing of an exoplanet's host star to direct and boost a signal to another solar system (Maccone, 2011; Hippke 2020).

Among the types of transmission technosignatures that can be searched for, one type that would be of heightened interest would be a wideband one (Messerschmitt, 2012 and Morrison, 2017) in the microwave range. Wideband conventionally has advantages over narrowband, such as less noise interference in tandem with more information capacity. And transmission in the microwave range, such as 30-90 GHz, would be an appealing option “for the designer of an interstellar beacon, based on efficiency considerations and taking account of the propagation characteristics” across the interstellar medium (Morrison, 2017).

## 2.2. ET Interplanetary Nav/Comm Networks

The development of an interplanetary nav/comm network by an ET civilization, especially across interstellar distances, would be possible for a civilization higher up the Kardashev scale than a civilization such as our own currently here on Earth (the Kardashev scale being a measure of a civilization's level of technological advancement based on the amount of energy it is able to use (Kardashev, 1964).

The manner in which an ET civilization would likely develop such a nav/comm network can be akin to the type of space exploration that the late physicist Freeman Dyson suggested (Giraud, 2016): A civilization can more effectively explore deep space by networking across it in incremental steps, such as across comets and rogue planets between the stars (versus passing up places in between and trying to do it all in a single trip). It is analogous to the proliferation of the Polynesians across the Pacific Ocean over several thousands of years, via their island-hopping by canoe. This is despite a civilization possibly being discouraged from traveling beyond its own solar system to the next star and beyond due to the unfathomably vast distances between the stars. Yet, amid our current civilization on Earth, there is noticeable support for the idea of sending nanoprobes propelled by lasers to our closest stellar neighbor, Alpha Centaur; an example is the Breaththrough Starshot initiative (Breakthrough Initiatives, 2020).

The interplay between staying close to home and venturing out, along with a need for efficiency based on the limitations of a civilization’s current technological innovation, may likely result in an interplanetary nav/comm network with a spoke-hub topology across interstellar distances. The spoke-hub topology is a natural topology in computer, business, and social networks found on modern-day Earth. A spoke-hub topology has several advantages over a mesh (or point-to-point) one with regard to efficiency. One advantage is that a spoke-hub network generally requires fewer routes; also, navigation or communication in a spoke-hub network travels shorter routes in a more centralized area. Much of the nav/comm activity would occur at the hub versus at a spoke connected to the hub.

This, along with the previous discussion in Section 2.1, lead to the first hypothesis of the modified approach:

> H1: A transmission technosignature would more likely be a wideband signal in the microwave or higher frequency range, which originates from the hub of an ET's interplanetary nav/comm network.

Note that each of the hypotheses presented herein are not necessarily dependent on one another from an analytical standpoint: Depending on the research goal of SETI researchers, the hypotheses can be analyzed individually, or together in tandem with one another.

### 2.3. X-Ray Pulsar-Based Navigation (XNAV)

An ET civilization capable of systematic interstellar nav/comm may incorporate into its interplanetary nav/comm network with an x-ray pulsar-based navigation (XNAV) system. XNAV serves as a galactic navigation satellite system for nav/comm by the ET civilization across interstellar distances through its interplanetary nav/comm network, whereby the XNAV's pulsars serve as beacons (Vidal, 2017). Thus, coupling of the XNAV system with the ET exoplanetary network represent a form of internetworking. Moreover, such internetworking can be characterized by the XNAV system having a center reference point in space that would be in common with the interplanetary nav/comm network, in terms of the network's hubs. This can be if the aforementioned ET civilization organizes these networks as such.

The potential of XNAV was demonstrated by NASA in 2017, via the space agency's NICER-SEXTANT mission aboard the International Space Station. NICER stands for Neutron star Interior Composition Explorer, while SEXTANT refers to Station Explorer for X-ray Timing and Navigation Technology. "The demonstration showed that millisecond pulsars could be used to accurately determine the location of an object moving at thousands of miles per hour in space—similar to how the Global Positioning System, widely known as GPS" functions (NASA, 2018).

Considering the implication that a potential ET exoplanetary network as described would likely entail an ET civilization capable of systematic interstellar nav/comm, it could behoove such a civilization to use XNAV as a galactic navigation positioning system, besides as a time standard. These would entail pulsars whose pulses would be observable and thus useable to the ET civilization.

This leads to the second hypothesis of the approach, which is:

> H2: A transmission technosignature would more likely be a wideband signal in the microwave or higher frequency range, which contains x-ray pulsar-based navigation (XNAV) metadata.

## 3. Method

The approach includes a method by which to test the two hypotheses mentioned in the previous section. Below is an overview of the method, followed by a case study that illustrates a preliminary application of the method.

## 3.1. Overview

As far as testing the first hypotheses, the method basically consists of conducting a network analysis of an exoplanetary database and pulsar database, whereby the results would prominently be displayed as 2-dimensional (2D) and 3-dimensionsal (3D) visualizations of hypothetical ET interplanetary nav/comm networks that would use XNAV. A Python software package known as a point processing toolkit (pptk), which is designed to process and visualize millions of data points, is used to facilitate this via a series of customized algorithms programmed into the pptk (Davis, 2019; Davis et al., 2020).

The following 9 steps outline the method in more detail relative to testing H1 and H2:

1. Create a network loading logic to read input data from an exoplanetary database (e.g., the NASA Exoplanet Archive). The logic includes converting celestial coordinates (right ascension and declination) into Euclidean ones (*x*, *y*, and *z*). Also, the network nodes would represent exoplanet host stars (host stars as proxies for their exoplanets when working with interstellar distances).
2. Plot the optimal nav/comm pathways between the nodes by experimenting with different "max distance" edge values, until reaching one that results in a network with a single component (no isolated, disconnected nodes); do this in conjunction with running Dijkstra's Algorithm. The resulting single component network would tend to resemble a natural spoke-hub network topology (versus a conventional mesh one), which would be similar to Earth-based computer, business, and social networks.
3. Visualize the network in 2D dimensionally preserving the *x* and *y* coordinate positions.
4. Do a community (or modular) analysis and colorize host stars based on their respective community ID (modularity class).
5. Do Degree, PageRank, HITS, and betweenness centrality analysis to find key star systems in a network, when modeling information propagation through the network. What makes the star systems key is that they represent key ”hubs” or “brokers” in the pathways of the network.
6. Visualize the network with ranked sizing of host stars based on the results of analytical scores, relative to the previous step (step 5).
7. 3D visualizations of the network preserving *x*, *y*, and *z* dimensions.
8. Identify a hub in the network, where a transmission technosignature is hypothetically more likely to be found.
9. Conduct a search for a wideband transmission technosignature within the hub that is wideband and in the microwave or higher frequency range, and that exhibits potential XNAV metadata.

The 9 steps above are a general set of steps in regard to the method, as the details of each step or the number of steps can potentially vary somewhat depending on the research context (the particular research objectives, instrumentation used, and so on). For example, in step 9, a researcher may also want to detect the same type of signal, but involving an exoplanetary pair in alignment or syzygy with Earth (where an ET would transmit an electromagnetic signal from one exoplanet in the pair to another, with Earth still somehow in line to detect the signal).

## 3.2. Case Study: NenuFAR SETI Project

One example of the method’s application, especially through network visualization, is a SETI project targeting 96 exoplanets for study using the French radio telescope NenuFAR; the 96 exoplanets are in NenuFAR’s field of view. A network analysis of the exoplanets was performed using data from the NASA Exoplanet Archive, with the results shown in Figure 3.1. Note that NenuFAR is optimized for the 10 MHz to 85 MHz frequency range, not the one for x-rays. To incorporate x-ray frequency data, NenuFAR would have to work in tandem with a near-Earth space satellite optimized for the x-ray frequency range, since x-rays do not readily penetrate the Earth’s atmosphere to reach its surface. Nevertheless, the application to NenuFAR illustrates the network visualization aspects of the method whether or not involving the x-ray frequency range.

**Figure 3.1.** A visualization of hypothetical ET interplanetary nav/comm network of 96 exoplanets targeted for observation using the French radio telescope NenuFAR, based on a community network analysis. (Figure courtesy of Caleb Jones)

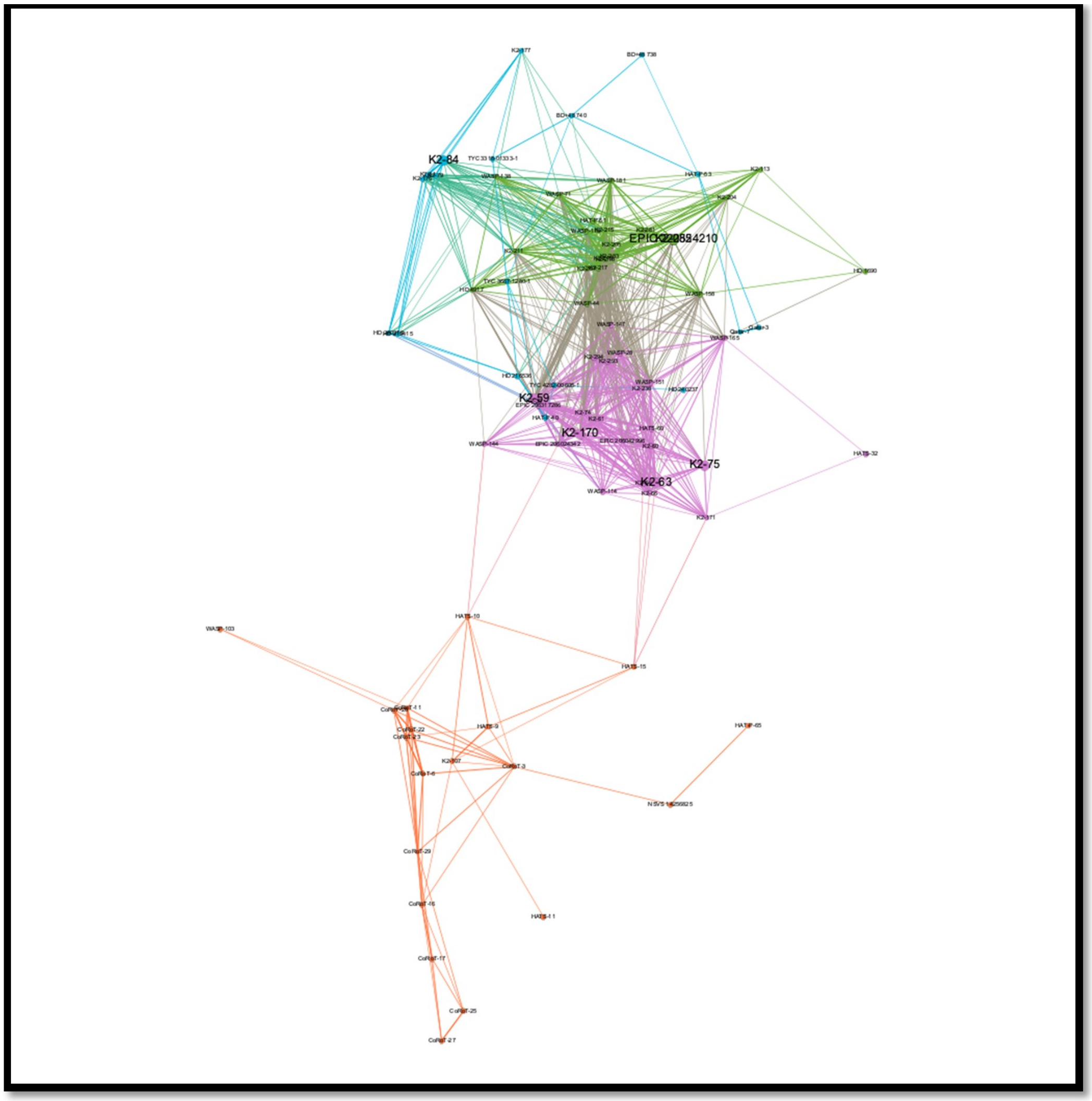

In the figure, each node in the network represents an exoplanet system's host star, while each edge or line represents a hypothetical nav/comm pathway among the nodes. The size of each node is determined by the number of confirmed or candidate exoplanets in the system: The more exoplanets there are in a system, the larger its node. The weight and length of each edge is determined using a single connected component set to a "max distance" of 300 parsecs, in conjunction with a type of network analysis known as modularity analysis. Edges of a specific color are associated with a local community of host stars that are more strongly connected to each other than to other stars in the wider community shown, i.e., the green edges represent one community, the purple edges represent another, and so on. Each of the local communities of host stars can be interpreted as subnetworks within the overall nav/comm network.

Note the higher concentration of nodes in the upper center part of the network shown in Figure 3.1, where the green, purple, and blueish colored subnetworks converge. This represents the "hub" of the overall network; a close-up of the hub is shown in Figure 3.2. The nodes are labeled according to the host star system they represent (e.g., K2-84 and K-81 in the upper left).

**Figure 3.2.** A close-up of the upper center portion or "hub" of the hypothetical ET interplanetary nav/comm network that was shown in Figure 3.1.

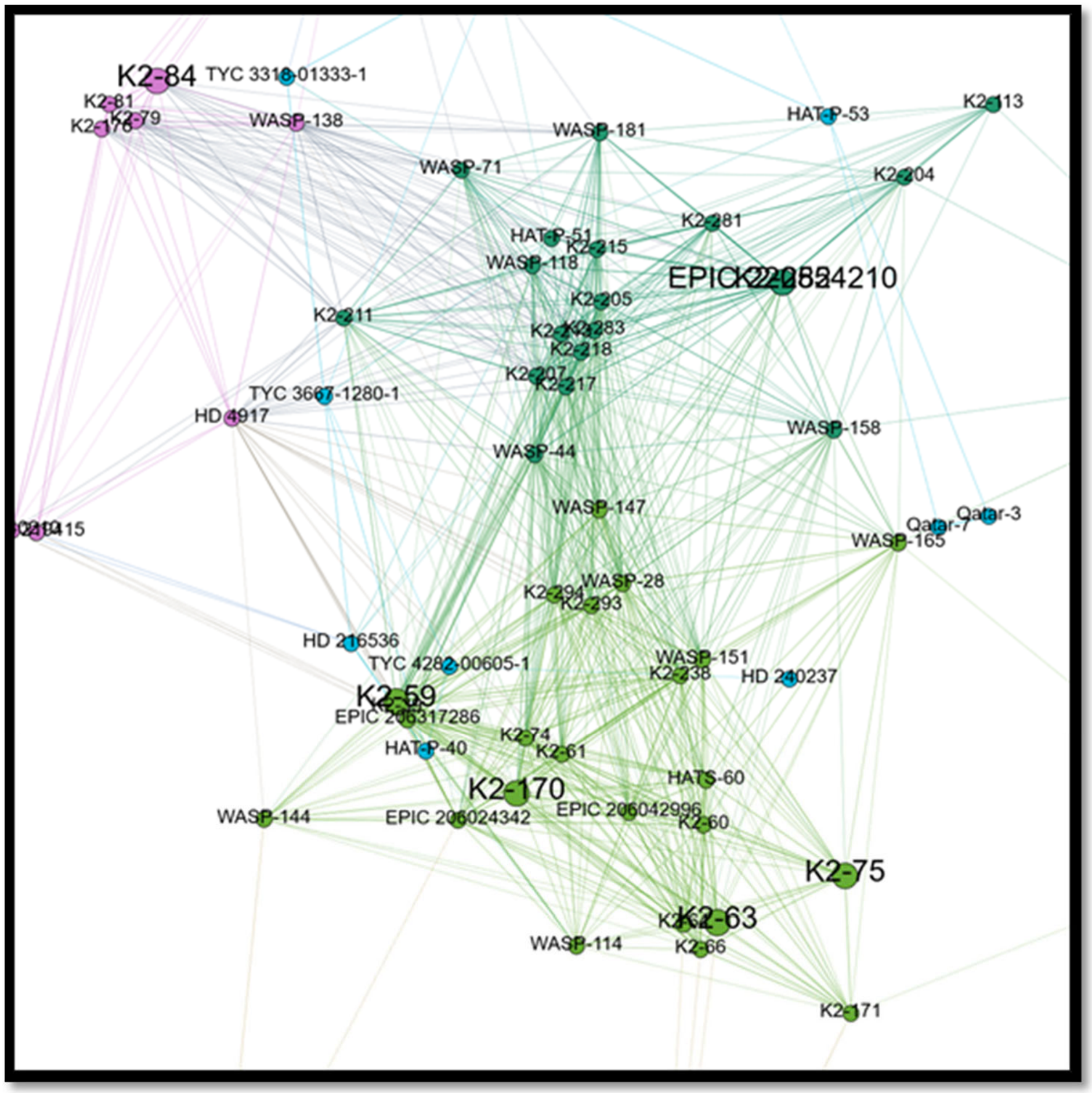

Figures 3.3 and 3.4 respectively depict how a communicative ET civilization could incorporate an XNAV system into its interplanetary nav/comm network, and how a transmission technosignature from such a network could appear to SETI researchers on Earth.

**Figure 3.3.** A model of three pulsars comprising an XNAV system, whose pulses an ET civilization can use within its interplanetary nav/comm network, e.g., to trilateralize in on the position of one of its spacecraft in the network hub, or for timing purposes therein. (Figure is a composite of Figure 3.1 courtesy of Caleb Jones and a 3-way pulsar diagram from Vidal, 2017)

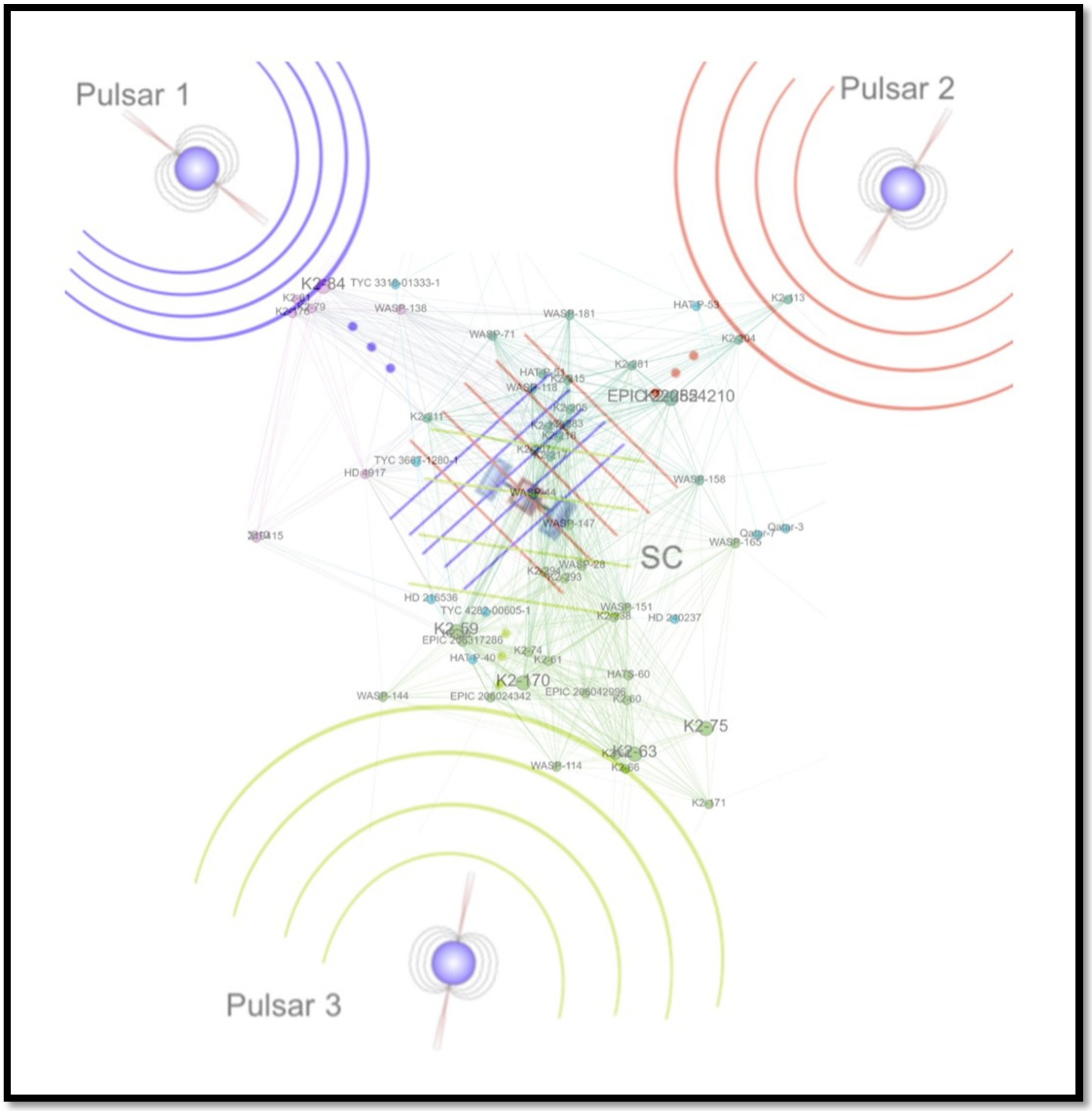

**Figure 3.4.** A hypothetical SETI signal broken down into metadata and data sections, with XNAV (x-ray pulsar-based navigation) and timing data in the metadata section.

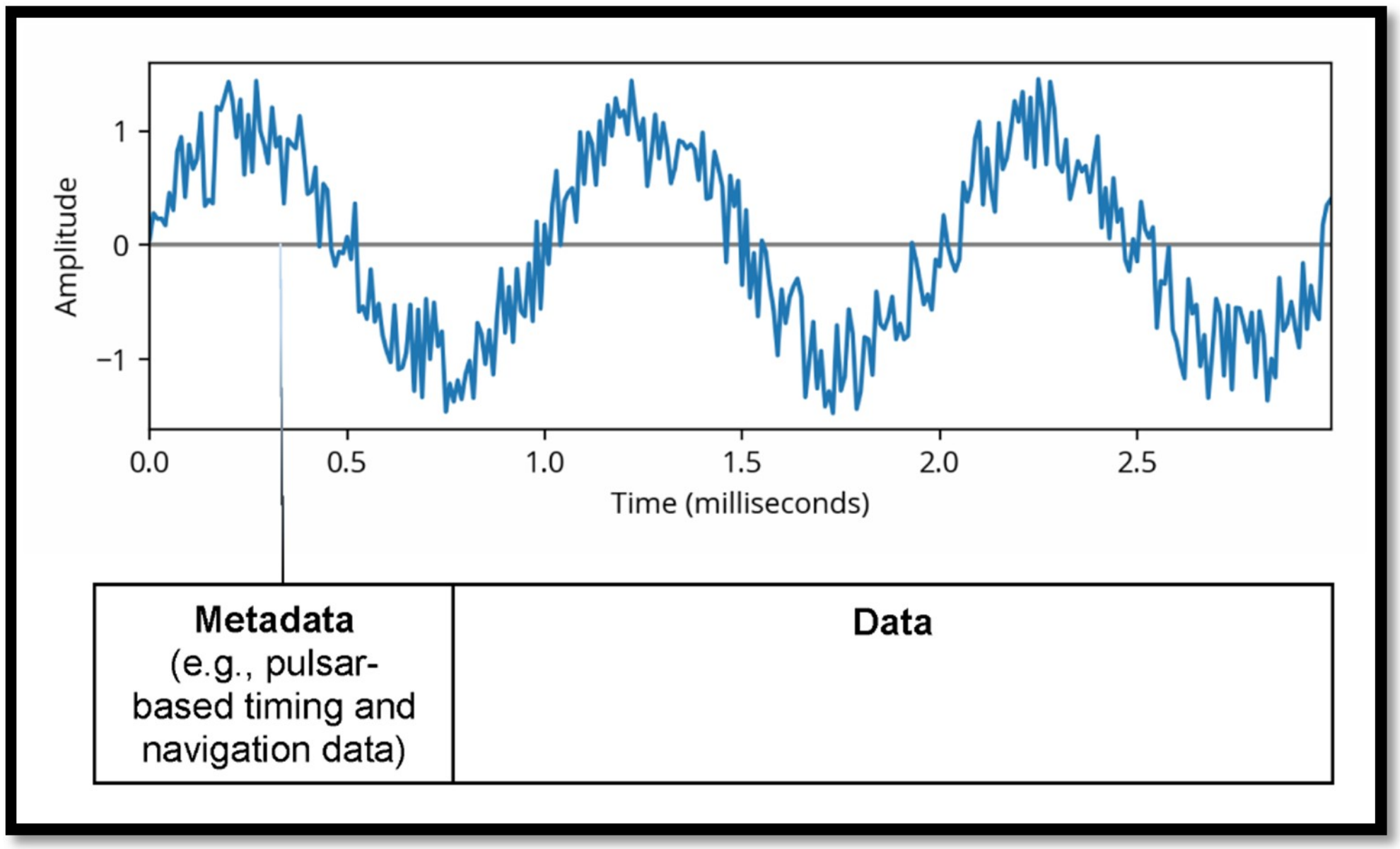

A significant possibility represented by Figure 3.4 is in the case of a transmission technosignature containing an apparent yet incomprehensible message, whether intended for us here on Earth and/or someone else in the cosmos. Even if we do not have some kind of alien language key to directly translate the message into something that we can understand, the overall form of the signal (the waveform) versus its content (the message) can nevertheless be an indirect means to interpret the message, by virtue of breaking the waveform down into smaller chunks or segments such as metadata and data, akin to the Internet protocol (IP) used here on Earth. What constitutes a given segment can be based on quantitative and/or qualitative research methods useful to the researcher. Each segment may be a clue to the type of information contained in the message, whereby further analysis can proceed from there in trying to determine what the overall message says, if it truly would say anything at all.

# 4. Conclusion

A transmission technosignature could be searched for through network analysis of a potential ET exoplanetary network with or without XNAV. Yet network analysis of an ET exoplanetary network coupled with XNAV offer the advantage of using multiple methods, so as to trilateralize in on and confirm a transmission technosignature.

Potential contributions to the field include 1) improved accuracy and efficiency in finding transmission technosignatures in the electromagnetic spectrum; 2) a common standard in reaching a Schelling Point (a mutual realization of how we and ETs can find each other); and 3) operationalizing the Drake Equation, esp. the $f_i$ term representing the fraction of planets with life that develop intelligent life.

Regarding the potential contribution of improved accuracy and efficiency in finding transmission technosignatures in the electromagnetic spectrum, this would be especially advantageous in fine-tuning targeted sky searches (besides all-sky searches) in terms of more accurately narrowing the search area, and thus the amount of data to analyze. And this can also save costs in terms of time, money, and other resources spent finding these technosignatures. Note that other technosignatures in the electromagnetic spectrum can include non-communicative ones, for example: unusual infrared emissions indicating waste heat from an ET technological process in an ET interplantary nav/comm network.

The increased accuracy in another sense increases the likelihood of reaching a Schelling Point, which is the point at which we and a communicative ET civilization both happen to come to the same realization about how we both should try to discover each other leading up to first contact. An example is a communicative ET civilization setting up a beacon to transmit a radio signal on a “magic frequency”—the frequency that the ET civilization hopes we will dial into; the signal may say something as simple as “We are here!” (Wright, 2018). Besides searching for the magic frequency, SETI researchers can draw upon the hypotheses to also consider where such a frequency would likely originate from in the cosmos.

The approach can potentially contribute to operationalizing the Drake Equation, by providing a means to help determine the fraction of planets in the Milky Way galaxy that contain communicative intelligent civilizations. The Drake Equation is shown below:

$$N = R_* f_p n_e f_l f_i f_c L$$

In the equation, N represents the number of communicative intelligent civilizations in the Milky Way galaxy that are detectable. $R_*$ represents the average rate of star formation in the Milky Way. $f_p$ is the fraction of stars in the Milky Way that have planets. $n_e$ is the average number of planets in the Milky Way that support life for each star that has planets. $f_l$ represents the fraction of planets in the Milky Way that can support life and where life would actually develop. $f_i$ represents the fraction of planets in the Milky Way that have life that develops into intelligent life. $f_c$ is the fraction such intelligent life that transmits detectable signals into space, that is, a communicative intelligent civilization. And L is the length of time for which such intelligent civilization transmits detectable signals into space (Drake, 1965).

Since the hypotheses draw upon concepts from an interdisciplinary perspective (such as network analysis used in information science, the social sciences, and the physical sciences), it can promote interdisciplinary research.

One potential limitation of the approach would be that it can be quite anthropomorphic, such as if an ET civilization develops nav/comm networks in space differently than what the hypotheses state. And an actual limitation is the limited exoplanetary data available as of this writing, so as to more fully plot interplanetary nav/comm networks: Only several thousand exoplanets have been confirmed thus far across limited portions of the sky (considering there is estimated to be billions of planets in the Milky Way galaxy, particularly Earth-sized ones)(NASA, 2020; Patigura et al., 2013; and Perlman, 2013). Nevertheless, the approach is a new opportunity that can inform future research whether in terms of the approach being useful or a stepping stone to another approach, so as to help facilitate one of the most important discoveries for humanity.

## Acknowledgements

Thanks to Clément Vidal for feedback about x-ray pulsar-based navigation (XNAV), and suggesting several scholarly sources; Greg Hellbourg for feedback about the NenuFAR SETI project; Ian Morrison for feedback about wideband space communications; and Caleb Jones for the exoplanet network visualization in Figure 3.1 based on exoplanetary data.

## Author Disclosure Statement

No competing financial interests exist.